\documentclass{article}
\usepackage{graphicx,amssymb, amsmath}
\title{Doped Spin Ladder: Zhang-Rice Singlets or Rung-centred Holes?}
\author{Krzysztof Wohlfeld\\
\scriptsize{Marian Smoluchowski Institute of Physics, Jagellonian University, 
Reymonta 4, PL-30059 Krak\'ow, Poland }}
\begin{document}
\date{}
\maketitle
\begin{abstract}
We formulate charge transfer model for a single doped spin ladder, relevant for the intrinsically
doped plane of coupled spin ladders in Sr$_{3}$Ca$_{11}$Cu$_{24}$O$_{41}$.
Assuming the presence of the experimentally observed charge order in the system 
we solve the model using Hartree-Fock approximation. Our results show
the profound stability of the Zhang-Rice singlets with respect to other 
configurations of doped holes, in agreement with recent x-ray absorption
measurements.
\end{abstract}
\section{INTRODUCTION}
It is a widely accepted feature of two-dimensional high temperature superconducting cuprates (HTSC) \cite{Sca06} that 
a hole doped into CuO$_2$ plane enters into a square of O ($2p$) orbitals and strongly binds with the 
central Cu ($3d$) hole forming a local, so-called Zhang-Rice \cite{Zha88} singlet (ZR; cf. Fig. \ref{fig:1}). 
On the other hand, a hole doped into the HTSC Cu$_2$O$_3$ coupled spin ladders can \textit{a priori} behave 
differently. This is due to the fact that: ($i$) spin ladders do not have the D$_{4h}$ symmetry (the four O 
surrounding Cu are not identical), and ($ii$) for {\it coupled} spin ladders 
the coordination number of the leg orbitals ($=3$) is different than that of the bridge orbital ($=2$) resulting in 
lower on-site energy for the hole in the bridge orbital than in one of the leg orbitals, cf. local-density
approximation studies in \cite{Mue98}. Hence, one could expect that e.g. a rung-centred state (rung) could be stabilized with a doped hole residing on the O ($2p$) bridge orbital and bound to the two neighbouring Cu holes via kinetic exchange interactions (cf. Fig. \ref{fig:1}). 

The aim of the present paper is to study the possible stability of \textit{these} two states in the doped spin ladder
by means of model calculations. It is an interesting theoretical task by itself, still backed by the recent x-ray absorption measurements which suggest that in the spin ladder plane of Sr$_{3}$Ca$_{11}$Cu$_{24}$O$_{41}$ (SCCO) a hole crystal with period $\lambda=3$ has been discovered with the doped holes evenly distributed among the O $2p$ orbitals involved in the ZR singlet \cite{Rus06}. This stays in contrast with the other experiments which suggest that the holes are not isotropically distributed. E.g. N\"ucker et al. \cite{Nue00} also in the x-ray absorption measurements find that the doped holes go mainly into those leg orbitals which are responsible for transport properties of the single ladder (crystalographic $c$ direction). Such result, however, not only contradicts those obtained in Ref. \cite{Rus06} but also a naive picture of rather localized holes in this {\it insulating} compound. Thus, the issue of the distribution of holes doped into spin ladders indeed needs some further studies. 
\section{MODEL HAMILTONIAN AND HARTREE-FOCK APPROXIMATION}
\label{sec:2}
\begin{figure}[t]
\includegraphics[width=\textwidth]{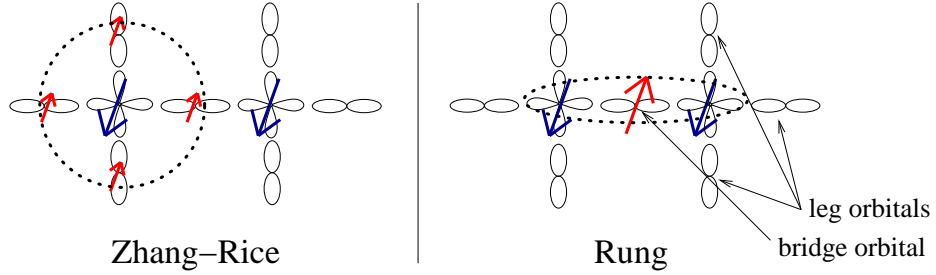}
\caption{\small{Schematic picture of the Zhang-Rice (ZR) singlet (left panel) and rung-centred (rung) hole (right panel)
in a Cu$_2$O$_7$ cluster. Large (small) arrows depict the hole spins for +1.0 (+0.25) charge. The grey arrows stand for spins of doped holes.}}
\label{fig:1}
\end{figure}
Our starting point is the three band charge transfer Hamiltonian relevant for a single doped spin ladder \cite{Woh06}:
\begin{align}
\label{eq:1}
H=&
\Big(-\!\!\!\sum_{m,j\in R,L,\sigma} 
     \!\!t_{mj} d^\dag_{m\sigma}p_{j\sigma}\!
-\!\!\!\sum_{m\in R,L;i\sigma} 
   \!\!t_{mi} d^\dag_{m\sigma}b_{i\sigma}+\mbox{h.c.}\!\Big) 
+\Delta\sum_{j\in R,L}n_{pj} + \varepsilon\sum_{i}n_{bi} \nonumber \\
&+ U\!\!\sum_{m\in R,L} n_{m\uparrow} n_{m\downarrow}           
+ U_p\Big(\sum_{i}       n_{bi\uparrow}n_{bi\downarrow} 
           +\sum_{j\in R,L}n_{pj\uparrow}n_{pj\downarrow}\Big),  
\end{align}
where: $\Delta$ is the charge transfer energy for the O ($2p_\sigma\equiv p$) {\it leg} orbital, 
$\varepsilon$ is the charge transfer for the O ($2p_\sigma\equiv b$) {\it bridge} orbital (cf. 
Fig. \ref{fig:1}), $t_{mj}$ is the hopping integral between the nearest neighbour pairs of the Cu 
($3d_{x^2-y^2}\equiv d$) orbital and $p$ or $b$ orbitals. We also include Coulomb on-site repulsion $U$ ($U_p$) 
between holes in the $d$ ($p$ or $b$) orbitals, respectively. Besides, the $d$ and $p$ orbitals 
belong to two sets of sites for the Left ($d_L,p_L \in \mbox{L}$) and Right ($d_R,p_R \in \mbox{R}$) 
leg of the ladder, and $n_m$, $n_{pj}$, $n_{bi}$ are particle number operators for $d$, $p$, $b$ 
orbitals, respectively.

In the following study we assume that the charge order is present in the ladder, as 
one of the few features of the ladder subsystem of SCCO
compound on which the experimentalists agree is the formation of the hole crystal \cite{Rus06}.
Therefore we expect that for commensurate values of doping a doped hole forms
a ZR or rung bound states which then form a hole crystal of relevant period. 

The presence of the charge order enables us to safely use Hartree-Fock approximation 
which works well for ordered states. This, however, means that we must adopt a
classical picture for ZR and rung states: a hole with the opposite $S^z$ component of 
the spin with respect to the spin of the relevant Cu holes involved in kinetic exchange
interactions (with no energy gain due to quantum fluctuations or phase coherence in ZR state). 
\section{RESULTS AND DISCUSSION}
\begin{figure}[t]
\includegraphics[width=\textwidth]{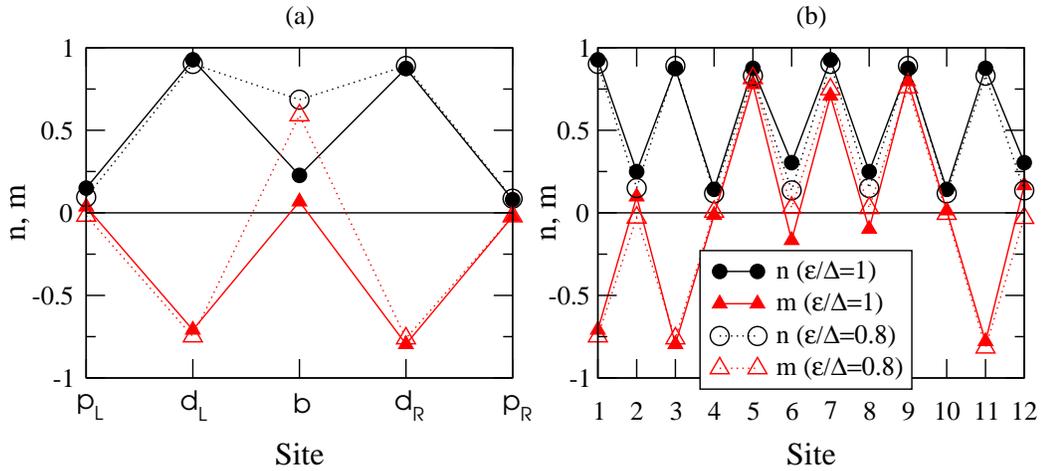}
\caption{\small{Hole density ($n$) and magnetization ($m$) for: (a) sites
along the rung with enhanced hole density; (b) 12 consecutive sites along one 
of the legs of the ladder (starting from the $d$ orbital, ie. for $d-p-d-...$
chain). The legend applies to both of the panels.}}
\label{fig:2}
\end{figure} 
We performed self-consistent numerical calculations using $60\times 7$ (the unit cell of the spin ladder consists of $7$
orbitals \cite{Woh06}) clusters for the following model parameteres: $U=9$ eV, $U_p=4.2$ eV, $\Delta=3.5$ eV \cite{Gra92}, 
whereas $t=1$ eV yields the correct superexchange constant $J/t\sim 0.1$, cf. Ref. \cite{Ecc98}.
We assume that the total number of holes per Cu $n_h=1.33$
in order to make our results comparable with the recent x-ray absorption measumerents in SCCO \cite{Rus06}.
Though, let us stress that the stability of ZR or rung states does not depend on the value
of the commensurate doping since similar results as shown below were obtained for other 
values of hole doping \cite{Woh06}. 

The results for two different values of the bridge orbital energy $\varepsilon$ are shown in Fig. \ref{fig:2}. 
For $\varepsilon/\Delta=1$ the hole distribution and magnetization resembles the classical ZR state: the doped hole is 
distributed rather isotropically among four O sites (e.g. sites $b$ and $p_L$ on Fig. \ref{fig:2}(a) 
and sites $2$ and $12$ on Fig. \ref{fig:2}(b)) surrounding the central Cu site 
occupied by roughly one hole (e.g. site $1$ on Fig. \ref{fig:2}(b) or $d_L$ on Fig. \ref{fig:2}(a)).
Also the spin of the doped hole in the O ($2p$) orbitals compensates roughly the spin of the hole 
in the Cu site. On the other hand, for $\varepsilon/\Delta=0.8$ the doped hole enters mainly into 
the $b$ orbital and there are much less holes in the $p$ orbitals of the leg
of the ladder suggesting a rung character of the doped hole. Let us also note that,
in agreement with the assumption of charge order, we find a hole crystal with less  
charge per every third rung of the ladder for the solution with ZR or rung character. 

\begin{figure}[t]
\includegraphics[width=\textwidth]{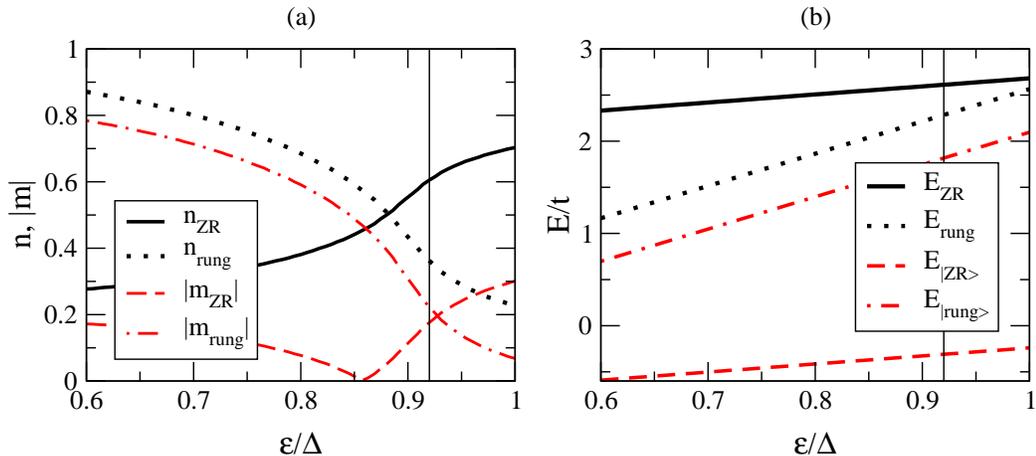}
\caption{\small{(a) Hole density ($n$) and absolute value of magnetization ($|m|$) 
for holes involved in forming ZR ($n_{\rm ZR}, |m_{\rm ZR}|$) and rung states
($n_{\rm rung}, |m_{\rm rung}|$) as a function of the of the bridge orbital energy
$\varepsilon$. (b) Binding energies of a single hole in ZR and rung states
in the classical case ($E_{ZR}$ and $E_{rung}$, respectively) and quantum-mechanical case ($E_{|ZR\rangle}$ 
and $E_{|rung\rangle}$, respectively) as a function of the the bridge orbital energy
$\varepsilon$. The vertical line on both panels depicts the value of 
$\varepsilon/\Delta=0.92$, cf. Ref. \cite{Mue98}.}}
\label{fig:3}
\end{figure}

In order to investigate the role of the specific spin ladder geometry on the stability of 
ZR and rung states we calculate the densities and magnetization of holes involved in forming:
($i$) ZR state: $n_{\rm ZR}=n_{p_L}+n_2+n_{12}$ and $|m_{ZR}|=|n_{ZR\uparrow}-n_{ZR\downarrow}|$
(we exclude the $b$ orbital from the sum to be able to distinguish between rung and ZR states), and 
($ii$) rung state: $n_{\rm rung}=n_b$ and $|m_{\rm rung}|=|n_{rung\uparrow}-n_{rung\downarrow}|$. 
The results are shown in Fig. \ref{fig:3}(a) as a function of the on-site energy of 
the bridge orbital $\varepsilon$. We find that with the decreasing value of $\varepsilon$
doped holes tend to occupy the $b$ orbital, and the spins of the holes in the $b$ orbital become 
polarized. Besides, the spins of the holes
involved in forming ZR state do not only compensate the spin of the central
Cu hole but for $\varepsilon/\Delta<0.85$ even weakly align ferromagnetically with the Cu spin.
Hence, we suggest that for $\varepsilon/\Delta<0.85$ the doped holes show a rung character 
while for $\varepsilon/\Delta>0.9$ they show a distinctive ZR character separated by a crossover regime. 
It means that ZR state is stable for the value of $\varepsilon/\Delta=0.92$, calculated in Ref. \cite{Mue98},
though we are very close to the crossover regime.

Let us now pose the question to what extent our results are relevant for the stability of the
real quantum-mechanical Zhang-Rice singlets or rung states. Therefore, using
{\it second} order perturbation theory in $U$ and $U-\Delta$ \cite{Zaa88} we calculate the 
binding energy of a single hole doped into ZR and rung
states: in the classical case ($E_{ZR}$ and $E_{rung}$, respectively), and
in the quantum-mechanical case ($E_{|ZR\rangle}$ and $E_{|rung\rangle}$, respectively).
In the classical case, which resembles the states obtained in the Hartree-Fock approximation,
we have:
\begin{equation}
E_{\rm ZR}=\frac{3}{4}\Delta+\frac{1}{4}\varepsilon+ J
\Big\langle \sum_{i\in ZR} \text{\bf{S}}_{i}\cdot \text{\bf{S}}_{\text{O}}\! -\! \frac{1}{4} \Big\rangle_{\rm ZR}
\enskip,    \enskip 
E_{\rm rung}=\varepsilon+J
\Big\langle \sum_{i\in rung}\text{\bf{S}}_{i} \cdot \text{\bf{S}}_{\text{O}}\! -\! \frac{1}{4} \Big\rangle_{\rm 
rung}
\end{equation}
where: the kinetic exchange $J=2t^2(1/U+1/(U-\Delta))$, $\text{\bf{S}}_{\text{O}}$ is the spin of the doped O ($2p$) 
hole, $\text{\bf{S}}_{i}$ is the spin of the Cu ($3d$) hole, and the sum goes over those Cu sites which 
are involved in forming a bound state with the O ($2p$) hole in rung or ZR state. 
The expressions for the energies in the quantum-mechanical case look similar except for
the averages of the spin operators which, unlike in the classical case, include also spin
fluctuations. In addition, for the real ZR singlet we include the phase coherence of holes
doped into the O ($2p$) orbitals \cite{Zha88}. The results as a function of the energy of the
bridge orbital $\varepsilon$ are shown in Fig. \ref{fig:3}(b). 

We find for the classical state that for $\varepsilon/\Delta<0.97$ the energy difference
($E_{ZR}-E_{rung}$) is larger than the effective hopping energy of the O ($2p$) hole ($=t^2/U$ or 
$t^2/(U-\Delta)$). Hence for finite bandwidth the rung state could only be stabilized up to 
the above value of the bridge orbital energy, qualitatively in agreement with the previous Hartree-Fock results.
However, in the quantum mechanical case the rung state could never be stabilized,
and due to the large energy difference ($E_{|ZR\rangle}-E_{|rung\rangle}$) the true Zhang-Rice singlet
should not be destabilized  by finite bandwidth.

\section{CONCLUDING REMARKS}
In summary, we have shown the profound stability of the Zhang-Rice singlets in the hole doped
spin ladders. First, using model Hartree-Fock calculations we obtain the isotropic 
distribution of doped holes among the O ($2p$) orbitals surrounding the central Cu ($3d$)
hole to be stable for the relevant range of the model parameters. Second, quantum-mechanical 
calculations of the binding energy of holes forming Zhang-Rice singlets and rung states 
suggest the Zhang-Rice singlets to be even more stable. Hence, our results
provide an interpretation of recent experimental data obtained in 
the x-ray absorption measumerents in the spin ladder plane of SCCO \cite{Rus06}.
 
\footnotesize{{\bf Acknowledgments}
I would like to thank the organising committee of the course for their financial support.
I am particularly grateful to Andrzej M. Ole\'s for his invaluable help, ideas, and comments 
and to George A. Sawatzky for very stimulating discussions. This work was supported by 
the Polish Ministry of Science and Education under Project No.~1 P03B 068 26.}


\begin{thebibliography}{99}
\bibitem{Sca06} D. J. Scalapino, arXiv:cond-mat/0610710 (unpublished).
                   [to appear in: "Numerical Studies of the 2D Hubbard Model",  
                   in \emph{Handbook of High Temperature Superconductivity}, 
                   edited by J. R. Schrieffer, Springer, 2006]

\bibitem{Zha88} F.C. Zhang and T.M. Rice, 
                   \emph{Phys. Rev. B} \textbf{37}, R3759 (1988).

\bibitem{Mue98} T. F. A. M\"uller et al., 
                   \emph{Phys. Rev. B} {\bf 57}, R12655 (1998).

\bibitem{Rus06}  A. Rusydi et al.,   
                   \emph{Phys. Rev. Lett.} {\bf 97}, 016403 (2006);
                  A. Rusydi et al.,
                   arXiv:cond-mat/0604101 (unpublished).

\bibitem{Nue00} N. N\"ucker et al.,
                   \emph{Phys. Rev. B} {\bf 62}, 14384 (2000).

\bibitem{Woh06} K. Wohlfeld, A. M. Ole\'s, and G. A. Sawatzky,  arXiv:cond-mat/0612669 (unpublished).
                   [to appear in Physica C: M2S-HTSC VIII Dresden 2006 Conference Proceedings, 2007]

\bibitem{Gra92} J.B. Grant and A.K. McMahan, 
                   \emph{Phys. Rev. B} {\bf46}, 8440 (1992).

\bibitem{Ecc98} R.S. Eccleston et al.,
                   \emph{Phys. Rev. Lett.} \textbf{81}, 1702 (1998).

\bibitem{Zaa88} J. Zaanen and A. M. Ole\'s, 
                   \emph{Phys. Rev. B} {\bf38}, 9423 (1988).

\end{thebibliography}
\end{document}